\documentclass[12pt]{article}
\input epsf.sty
\def\be{\begin{equation}}
\def\ee{\end{equation}}
\def\ba{\begin{align}}
\def\ea{\end{align}}

\def\p{\partial}

\def\ops[#1]{\p_{#1} e^{-2\phi}}

\def\eq[#1]{equation (\ref {eq:#1})}
\def\Eq[#1]{Equation (\ref {eq:#1})}
\def\e[#1]{\ref {eq:#1}}
\def\at[#1]{| _{#1}}

\let\oldpercent\%\renewcommand{\%}{\scalebox{0.85}{\oldpercent}}

\begin{document}

\baselineskip=18pt

\begin{center}
{\Large \bf{Explicit microstates at the Schwarzschild horizon}}

\vspace{10mm}

\textit{ Amit Giveon}
\break

Racah Institute of Physics, The Hebrew University \\
Jerusalem, 91904, Israel

\end{center}

%\break

\vspace{10mm}

\begin{abstract}

The claim that the microstates of Schwarzschild black holes in perturbative string theory
amount to the modes of long folded strings in the vicinity of its horizon is supported by more evidence.

\end{abstract}
\vspace{10mm}

In this short note, we argue that in perturbative string theory, the microstates that account for the Schwarzschild black hole (BH) entropy
are localized at the vicinity of the horizon, and suggest what they look like.

Concretely, let~\cite{Horowitz:1997jc}
\be\label{iii}
I=I_{gravity}+I_W
\ee
be a Euclidean action in $D=1+d$ dimensional spacetime
with a spatial metric, $g_{ij}$, $i,j=1,\dots,d$, a thermal circle with component
\be\label{gphi}
g_{00}\equiv e^{2\phi}~,
\ee
a $d$-dimensional dilaton, $\Phi_d$, reduced from the $D$-dimensional one by
\be\label{dilaton}
\Phi_d\equiv \Phi_D-{1\over 4}\log g_{00}=\Phi_D-{\phi\over 2}~,
\ee
and a winding string condensate, $W^\pm$.

The first term in (\ref{iii}) is the pure gravity-dilaton one,
\be\label{igravity}
I_{gravity}=-{\beta\over 16\pi G_N}\int d^dx\sqrt{g_D}e^{-2\Phi_D}\left({\cal R}_d+4g^{ij}_d\partial_i\Phi_d\partial_j\Phi_d-g^{ij}_d\partial_i\phi\partial_j\phi\right)~,
\ee
while the second term is the effective action of the winding string condensate,
\be\label{iw}
I_W={\beta\over 16\pi G_N}\int d^dx\sqrt{g_D}e^{-2\Phi_D}\left(g^{ij}_d\partial_i W^+\partial_j W^-+m^2_W(g_{00})W^+W^-\right)~,
\ee
where the $\phi$-dependent mass of the thermal scalar is
\be\label{mmbb}
m^2_W(\phi)={1\over(2\pi\alpha')^2}\left(-\beta_H^2+\beta^2g_{00}(\phi)\right)~,
\ee
with ${1\over 2\pi\alpha'}$ being the string tension, which sets in particular the string length scale, $\ell_s\equiv\sqrt{\alpha'}$.
The first term in (\ref{mmbb}) is the mass squared of a closed string tachyon
and the second is the contribution due to the length of a string winding once around the thermal circle
(whose local radius is $\sqrt{g_{00}}=e^\phi$, (\ref{gphi})).

The action (\ref{iii})--(\ref{mmbb})
is cooked such that on a Schwarzschild BH background the Euclidean action of the winding condensate vanishes
on its solution, $I_W=0$, as it must.~\footnote{String theory is cooking the effective action automatically this way:
In string theory on the $SL(2)/U(1)$ cigar, we know that $W$ is an exact zero-mode, hence,
whatever the effective theory is, it should be constructed such that, in particular, it satisfies this property.
The idea is thus to write down an effective action for a thermal scalar which is an exact zero-mode on generic Euclidean black holes.
The action~(\ref{iii})--(\ref{mmbb}) has this property, so it is a good starting point.
Now, in string theory there are of course corrections, however, for parametrically small string coupling and
parametrically small curvature, it is reasonable to expect that (\ref{iii})--(\ref{mmbb}) is a good approximation.}
This implies that on the e.o.m. one necessarily has $I=\beta F$, where $F$ is the free energy of the BH and $\beta$ its inverse temperature;
consequently, the entropy,
\be\label{entropy}
S\equiv (\beta\partial_\beta-1)I=S_{BH}~,
\ee
is the BH entropy, by construction!

An interesting point of \cite{Chen:2021emg,Brustein:2021cza} is that, on the other hand,
\be\label{point}
S\equiv (\beta\partial_\beta-1)I={\beta\over 16\pi G_N}\int d^dx\sqrt{g_D}e^{-2\Phi_D}{2\beta^2e^{2\phi}\over(2\pi\alpha')^2}W^+W^-~,
\ee
which has remarkable consequences, following
\cite{Kutasov:2005rr,Giveon:2012kp,Giveon:2013ica,Mertens:2013zya,Giveon:2014hfa,Mertens:2015hia,Giveon:2019gfk,Giveon:2019twx,Giveon:2020xxh,Jafferis:2021ywg},
both technical and conceptual. Concretely, for large BHs,
the winding condensate profile near the horizon is sufficient to obtain the entire BH entropy.

Indeed, it can be verified that in the $R^2$ limit near the horizon, a.k.a.
the regime where the $D$-dimensional metric is approximately
$ds^2=d\rho^2+\rho^2d\theta^2+r_0^{D-2}d\Omega_{D-2}^2$
with $\rho$ being the invariant distance from the tip of the Euclidean BH cigar,
plugging $\phi=\log\rho$, (\ref{gphi}), and
\be\label{www}
W^+W^-=e^{-{\rho^2\over \alpha'}}
\ee
in (\ref{point}),
where (\ref{www}) is the (square of the) winding string zero-mode profile in the $R^2$ limit~\cite{Giveon:2012kp}, gives
\be\label{sss}
S={A_H\over 4G_N}~,
\ee
where $A_H$ is the area of the horizon. The Gaussian decay of the condensate, (\ref{www}),
guaranties the high validity of the approximation
when the Schwarzschild radius is sufficiently bigger then the string length scale, $r_0\gg\ell_s$.

The conceptual consequences are far reaching.
Long ago, starting in
\cite{Giveon:2012kp,Giveon:2013ica,Mertens:2013zya,Giveon:2014hfa}, the following question was addressed:
``Is the tip of the cigar geometry special in the small curvature limit?''
It was argued that the answer is positive,~\footnote{For instance, it was shown in
\cite{Giveon:2012kp,Giveon:2013ica,Mertens:2013zya,Giveon:2014hfa} that
``the contribution of the discrete states is present even for perturbative strings propagating in the background of large Schwarzschild black
holes; it was argued that the discrete states live at a stringy distance from the tip of the cigar both from the conformal field theory wave-function analysis and other perspectives, thus, the way string theory takes care of its self-consistency seems to have important consequences for the physics near horizons, even for parametrically large black holes.''}
eventhough naively one might expect the physics of regular $R^2$, instead;
the result (\ref{point}) is of course supporting this surprising claim.~\footnote{To conclude that (\ref{point}) carries the whole entropy,
it is implicitly assumed here that the tip of the cigar is a special point
such that, in particular, (effectively) the Euclidean circle doesn't shrink.
And, as mentioned above, in string theory on the cigar, this may indeed be the case.}

To identify the microstates at the horizon we make the following {\it assumptions:}
\begin{enumerate}
\item
For our purposes below -- estimating the number of explicit microstates at the horizon of a large Schwarzschild BH and see what they look like,
it is sufficient to approximate the near-horizon theory of a $D$-dimensional Schwarzschild BH with temperature $1/\beta$
by an $SL(2)_k/U(1)$ exact SCFT,~\footnote{We consider BHs in the type II superstring.} with the level $k$ related to $\beta$ via~\footnote{From
the value of the $2d$ string coupling at the tip (a.k.a. horizon), $g_s$, and the area of the horizon $D$-sphere, $S^{D-2}$,
one can then have also the $D$-dimensional Newton constant, $G_N$,
and consequently, from $\beta$ and $G_N$, also the BH mass and entropy, $M$ and $S$ (in the standard way),
expressed in terms of the string length scale, $\ell_s$, and the parameters, $g_s,k,D$.}
\be\label{kkk}
\alpha' k=\left({2r_0\over D-3}\right)^2=\left({\beta\over 2\pi}\right)^2~.
\ee
This assumption has strong support in \cite{Chen:2021emg}, following \cite{Emparan:2013xia}.~\footnote{The estimations
below are for parametrically large Schwarzschild radii, $r_0/\ell_s\gg 1$,
and in the $R^2\times S^{D-2}$ approximation near the tip (horizon)
it is reasonable to assume that the physics is that of the near-horizon
$SL(2)_k/U(1)\times S^{D-2}$ ``almost CFT''~\cite{Chen:2021emg,Emparan:2013xia}, at large $k$, (\ref{kkk}).}
\item
The number $N$ of ``string bits'' which amounts to the condensate~\footnote{We are schematic here;
see \cite{Giveon:2019gfk,Giveon:2019twx} for details.}
$F\sim W^+W^-$ in the $2d$ $SL(2)_k/U(1)$ BH is (see \cite{Giveon:2020xxh} for details)
\be\label{nnn}
N=kN_{IFS}=2\pi e^{-2\Phi_{hor}}~,
\ee
where $\Phi_{hor}$ is the value of the two-dimensional dilaton on the horizon and $N_{IFS}$ stands for the number of
`instantly created folded strings' (IFSs) of~\cite{Itzhaki:2018glf,Attali:2018goq}
at the vicinity of the BH horizon. This assumption has strong support in~\cite{Giveon:2020xxh} and below.~\footnote{The 
number of IFSs that fill the eternal $SL(2)_k/U(1)$ black hole was calculated in~\cite{Giveon:2020xxh}
{\it indirectly}, subject to some assumptions, in three different ways that gave the exact same answer:
Since the trigger for the IFS creation is a time-like dilaton,
a natural way to determine their number is to ask how many IFSs
are needed for their backreaction to render the dilaton time-independent behind the horizon.
Moreover, the IFSs violate the ANEC~\cite{Attali:2018goq}, and so it is possible that a sufficient number of
them can prevent particles from falling into the black hole. It turns out that~(\ref{nnn}) is
exactly that number. The third way that gives~(\ref{nnn}) in~\cite{Giveon:2020xxh} is via entropy considerations, 
again, subject to assumptions.}

\end{enumerate}

%\noindent
{\it Consequence:}
\be\label{sn}
S=N~,
\ee
where $S$ and $N$ are those in (\ref{point}),(\ref{sss}) and (\ref{nnn}), respectively;
a.k.a, the microstates which give rise to the BH entropy correspond to the long strings modes at the (stringy) vicinity of its horizon.

%\noindent
{{\it Comment:}

We know that the assumptions are correct for large $D$, $\beta/\ell_s$ and $k$;
needless to say that if they are modified in other regimes of the $(D,\beta/\ell_s,k)$ parameters space~\footnote{The
size of $\beta/\ell_s$ and $k$ is not independent, (\ref{kkk});
assumption 1 is correct also for small $k$ and $\beta/\ell_s$.}
then the detailed consequences should be reexamined accordingly.~\footnote{For instance, we know \cite{Giveon:2005mi} that for $k<1$, a.k.a. below the BH/string transition, the $k$ dependence of $S(M)$ is modified (even for parametrically large $D$); it is possible that the
``fractionation of the long IFS to $k$ little strings,'' (\ref{nnn}), which is argued at large $k$, is modified accordingly at $k<1$.}

To support (\ref{nnn}), a.k.a. that the IFS
can be thought of as a ``fractionated fundamental string, bound to the BH horizon, with inverse tension $2\pi\alpha'/k$,''
we consider a classical folded string in $R_\phi\times S^1_X$,
where the real scalar $\phi$ has a linear dilaton with a slope $Q$,
such that the string coupling is
\be\label{gsq}
g_s\equiv e^\Phi=e^{-Q\phi/\alpha'}~,
\ee
and the scalar $X$ is compact with a radius $R_X$,
which is related to the dilaton slope and to the level $k$ in (\ref{kkk}),(\ref{nnn}) by
\be\label{rkq}
R_X=\sqrt{\alpha' k}=\alpha'/Q~.
\ee
In what follows, the analysis is strictly valid for large integer $k$.

The folded string solution~\cite{to appear},
obtained by analytic continuation of that in \cite{Maldacena:2005hi}, is
\be\label{folded}
X=\ell_s\sigma_1~,\qquad \phi=Q\log\left(\cos(\ell_s\sigma_1/Q)+\cosh(\ell_s\sigma_2/Q)\right)~.
\ee
At $\sigma_2\to\pm\infty$, the radial field $\phi$ blows up at weak coupling; in this regime, (\ref{folded}) looks like
a pair of incoming strings, winding in opposite orientations around the $X$ direction; we identify them with $W^\pm$.
At $\sigma_2=0$ and $\sigma_1=Q(2\pi n-\pi)/\ell_s$, with $n=1,...,k$, the radial
$\phi$ blows up at strong coupling, instead; in this regime, (\ref{folded}) looks like $k$ little strings.
The two asymptotic behaviors of (\ref{folded}) are connected via string folds at $\phi=0$.

After turning on an $N=2$ Liouville wall, the above describes aspects of the physics of a folded string in the cigar CFT,~\footnote{In
the bosonic case, one turns on a sine-Liouville wall, instead; see~\cite{Giveon:2015cma} and references therein/thereof for more aspects
of the cigar/sine-Liouville ($N=2$ Liouville) correspondence.}
and upon continuation to real time, it describes aspects of the physics of an instantly created long
(of order $\sqrt{k}\ell_s\sim r_0$, (\ref{kkk})) folded string
in the vicinity (of order $\ell_s$, (\ref{www})) of an $SL(2)_k/U(1)$ BH horizon~\cite{to appear}.
Finally, following the assumptions above, (\ref{kkk}),(\ref{nnn}),
this also gives rise to the number of bits carried by each one of the $N_{IFS}$ long strings
at the horizon of a large Schwarzschild black hole, altogether giving rise to its entropy, (\ref{sn}),(\ref{sss}).

To recapitulate, within the assumptions above,
we identified the microstates of large Schwarzschild black holes in perturbative string theory.
Needless to say that a lot more awaits to be understood; we hope to report on that in the future.

\vspace{10mm}

\section*{Acknowledgments}
I thank N.~Itzhaki for collaboration on many parts of this letter.
This work is supported in part by a center of excellence supported by the Israel Science
Foundation (grant number 2289/18) and BSF (grant number 2018068).


\vspace{10mm}


\begin{thebibliography}{1}

%\cite{Horowitz:1997jc}
\bibitem{Horowitz:1997jc}
G.~T.~Horowitz and J.~Polchinski,
``Selfgravitating fundamental strings,''
Phys. Rev. D \textbf{57}, 2557-2563 (1998)
doi:10.1103/PhysRevD.57.2557
[arXiv:hep-th/9707170 [hep-th]].
%155 citations counted in INSPIRE as of 08 Aug 2021

%\cite{Chen:2021emg}
\bibitem{Chen:2021emg}
Y.~Chen and J.~Maldacena,
``String scale black holes at large $D$,''
[arXiv:2106.02169 [hep-th]].
%4 citations counted in INSPIRE as of 05 Aug 2021

%\cite{Brustein:2021cza}
\bibitem{Brustein:2021cza}
R.~Brustein and Y.~Zigdon,
``Black Hole Entropy Sourced by String Winding Condensate,''
[arXiv:2107.09001 [hep-th]].
%0 citations counted in INSPIRE as of 05 Aug 2021

%\cite{Kutasov:2005rr}
\bibitem{Kutasov:2005rr}
D.~Kutasov,
``Accelerating branes and the string/black hole transition,''
[arXiv:hep-th/0509170 [hep-th]].
%57 citations counted in INSPIRE as of 06 Aug 2021

%\cite{Giveon:2012kp}
\bibitem{Giveon:2012kp}
A.~Giveon and N.~Itzhaki,
``String Theory Versus Black Hole Complementarity,''
JHEP \textbf{12}, 094 (2012)
doi:10.1007/JHEP12(2012)094
[arXiv:1208.3930 [hep-th]].
%63 citations counted in INSPIRE as of 05 Aug 2021

%\cite{Giveon:2013ica}
\bibitem{Giveon:2013ica}
A.~Giveon and N.~Itzhaki,
``String theory at the tip of the cigar,''
JHEP \textbf{09}, 079 (2013)
doi:10.1007/JHEP09(2013)079
[arXiv:1305.4799 [hep-th]].
%39 citations counted in INSPIRE as of 05 Aug 2021

%\cite{Mertens:2013zya}
\bibitem{Mertens:2013zya}
T.~G.~Mertens, H.~Verschelde and V.~I.~Zakharov,
``Random Walks in Rindler Spacetime and String Theory at the Tip of the Cigar,''
JHEP \textbf{03}, 086 (2014)
doi:10.1007/JHEP03(2014)086
[arXiv:1307.3491 [hep-th]].
%28 citations counted in INSPIRE as of 05 Aug 2021

%\cite{Giveon:2014hfa}
\bibitem{Giveon:2014hfa}
A.~Giveon, N.~Itzhaki and J.~Troost,
``Lessons on Black Holes from the Elliptic Genus,''
JHEP \textbf{04}, 160 (2014)
doi:10.1007/JHEP04(2014)160
[arXiv:1401.3104 [hep-th]].
%25 citations counted in INSPIRE as of 05 Aug 2021

%\cite{Mertens:2015hia}
\bibitem{Mertens:2015hia}
T.~G.~Mertens, H.~Verschelde and V.~I.~Zakharov,
``The long string at the stretched horizon and the entropy of large non-extremal black holes,''
JHEP \textbf{02}, 041 (2016)
doi:10.1007/JHEP02(2016)041
[arXiv:1505.04025 [hep-th]].
%19 citations counted in INSPIRE as of 05 Aug 2021

%\cite{Giveon:2019gfk}
\bibitem{Giveon:2019gfk}
A.~Giveon and N.~Itzhaki,
``Stringy Black Hole Interiors,''
JHEP \textbf{11}, 014 (2019)
doi:10.1007/JHEP11(2019)014
[arXiv:1908.05000 [hep-th]].
%10 citations counted in INSPIRE as of 05 Aug 2021

%\cite{Giveon:2019twx}
\bibitem{Giveon:2019twx}
A.~Giveon and N.~Itzhaki,
``Stringy Information and Black Holes,''
JHEP \textbf{06}, 117 (2020)
doi:10.1007/JHEP06(2020)117
[arXiv:1912.06538 [hep-th]].
%10 citations counted in INSPIRE as of 05 Aug 2021

%\cite{Giveon:2020xxh}
\bibitem{Giveon:2020xxh}
A.~Giveon, N.~Itzhaki and U.~Peleg,
``Instant Folded Strings and Black Fivebranes,''
JHEP \textbf{08}, 020 (2020)
doi:10.1007/JHEP08(2020)020
[arXiv:2004.06143 [hep-th]].
%4 citations counted in INSPIRE as of 05 Aug 2021

%\cite{Jafferis:2021ywg}
\bibitem{Jafferis:2021ywg}
D.~L.~Jafferis and E.~Schneider,
``Stringy ER=EPR,''
[arXiv:2104.07233 [hep-th]].
%5 citations counted in INSPIRE as of 05 Aug 2021

%\cite{Emparan:2013xia}
\bibitem{Emparan:2013xia}
R.~Emparan, D.~Grumiller and K.~Tanabe,
``Large-D gravity and low-D strings,''
Phys. Rev. Lett. \textbf{110}, no.25, 251102 (2013)
doi:10.1103/PhysRevLett.110.251102
[arXiv:1303.1995 [hep-th]].
%82 citations counted in INSPIRE as of 05 Aug 2021

%\cite{Itzhaki:2018glf}
\bibitem{Itzhaki:2018glf}
N.~Itzhaki,
``Stringy instability inside the black hole,''
JHEP \textbf{10}, 145 (2018)
doi:10.1007/JHEP10(2018)145
[arXiv:1808.02259 [hep-th]].
%9 citations counted in INSPIRE as of 11 Aug 2021

%\cite{Attali:2018goq}
\bibitem{Attali:2018goq}
K.~Attali and N.~Itzhaki,
``The Averaged Null Energy Condition and the Black Hole Interior in String Theory,''
Nucl. Phys. B \textbf{943}, 114631 (2019)
doi:10.1016/j.nuclphysb.2019.114631
[arXiv:1811.12117 [hep-th]].
%8 citations counted in INSPIRE as of 11 Aug 2021

%\cite{Giveon:2005mi}
\bibitem{Giveon:2005mi}
A.~Giveon, D.~Kutasov, E.~Rabinovici and A.~Sever,
``Phases of quantum gravity in AdS(3) and linear dilaton backgrounds,''
Nucl. Phys. B \textbf{719}, 3-34 (2005)
doi:10.1016/j.nuclphysb.2005.04.015
[arXiv:hep-th/0503121 [hep-th]].
%86 citations counted in INSPIRE as of 05 Aug 2021


\bibitem{to appear}
A.~Hashimoto, N.~Itzhaki and U.~Peleg, Unpublished work.

%\cite{Maldacena:2005hi}
\bibitem{Maldacena:2005hi}
J.~M.~Maldacena,
``Long strings in two dimensional string theory and non-singlets in the matrix model,''
JHEP \textbf{09}, 078 (2005)
doi:10.1088/1126-6708/2005/09/078
[arXiv:hep-th/0503112 [hep-th]].
%66 citations counted in INSPIRE as of 05 Aug 2021

%\cite{Giveon:2015cma}
\bibitem{Giveon:2015cma}
A.~Giveon, N.~Itzhaki and D.~Kutasov,
``Stringy Horizons,''
JHEP \textbf{06}, 064 (2015)
doi:10.1007/JHEP06(2015)064
[arXiv:1502.03633 [hep-th]].
%35 citations counted in INSPIRE as of 07 Aug 2021


\end{thebibliography}
\end{document}